\documentclass{ws-procs9x6}
\usepackage{graphicx}
\newcommand{\half}{\frac{1}{2}}
\newcommand{\SW}{Sheikholeslami--Wohlert}

\newcommand{\pslash}{\not\!p}

\newcommand{\gm}{\gamma_{\mu}}

\newcommand{\cqqg}{\cite{Skullerud:2002ge}}
\newcommand{\cbc}{\cite{Ball:1980ay}}
\begin{document}

\title{The nonperturbative quark--gluon 
vertex\footnote{\uppercase{P}resented by \uppercase{J}.~\uppercase{S}kullerud}}

\author{Jonivar Skullerud\footnote{\uppercase{W}ork 
supported by \uppercase{F}\uppercase{O}\uppercase{M}.}} 

\address{Instituut for Theoretische Fysica, University of Amsterdam\\
Valckenierstraat 65, 1018 XE Amsterdam, The Netherlands\\
Email: jonivar@skullerud.name}

\author{Patrick Bowman and Ay{\c s}e Kizilers\"u}

\address{Centre for the Subatomic Structure of Matter, University of
Adelaide\\Adelaide, SA 5005, Australia}

\maketitle

\abstracts{
We show results for the quark--gluon vertex in the Landau gauge, using
a mean-field improved Sheikholeslami--Wohlert fermion action.  We
compute all the three non-zero form factors of the vertex at zero
gluon momentum, and compare them to the abelian vertex.  The quark
mass dependence of the vertex is also investigated and found to be
negligible for the range of masses considered.}

\section{Introduction}
\label{sec:intro}

The quark--gluon vertex plays an important role in the dynamics of
QCD.  A non-trivial infrared structure of the vertex can affect
hadronic observables through the running of the relevant coupling,
while at a more fundamental level it is thought crucial to achieve
confinement and a sufficient degree of dynamical chiral symmetry
breaking.  Therefore, a first-principles, nonperturbative
determination of the vertex, which the lattice may provide, is highly
desirable. 

In a previous paper\cqqg, results for the form factor ($\lambda_1$)
proportional to the running coupling were presented.  Here, we extend
this by also studying the two other nonzero form factors at zero gluon
momentum.  We also study the mass dependence of these form factors.

We use the notation and the decomposition of the vertex given in
Ref\cqqg.  The outgoing quark momentum is denoted $p$ and the (outgoing)
gluon momentum $q$.  The incoming quark momentum is thus $k=p+q$.  For
$q=0$, the vertex can be written
\begin{equation}
\Lambda_\mu(p,q) = -ig\Bigl[\lambda_1(p^2)\gm - 4\lambda_2(p^2)\pslash p_\mu 
 -2i\lambda_3(p^2)p_\mu\Bigr] \, .
\label{eq:bc}
\end{equation}
In an abelian theory (QED), the Ward--Takahashi identities mean that
these form factors are given uniquely in terms of the fermion
propagator $S^{-1}(p)=i\pslash A(p^2)+B(p^2)$ by\cbc
\begin{equation}
\lambda_1^{\mathrm{QED}}(p^2) = A(p^2) \qquad
\lambda_2^{\mathrm{QED}} = -\half\frac{\mathrm{d}}{\mathrm{d}p^2}A(p^2) \qquad
\lambda_3^{\mathrm{QED}} = -\frac{\mathrm{d}}{\mathrm{d}p^2}B(p^2)
\end{equation}
By comparing with these expressions, we can thus get a direct measure
of the non-abelian nature of the vertex.

All the results presented here are obtained in the Landau gauge, in
the quenched approximation at $\beta=6.0$, using the \SW\ fermion
action with mean-field improvement coefficients.  We have used two
quark masses, $\kappa=0.137$ and 0.1381, corresponding to a bare quark mass
$m=118$ and 61 MeV respectively, with
about 500 configurations in both cases.  For further details about the
calculation, we refer to Ref\cqqg.

\section{Results}
\label{sec:results}

\begin{figure}[ht]
\includegraphics*[width=0.49\textwidth]{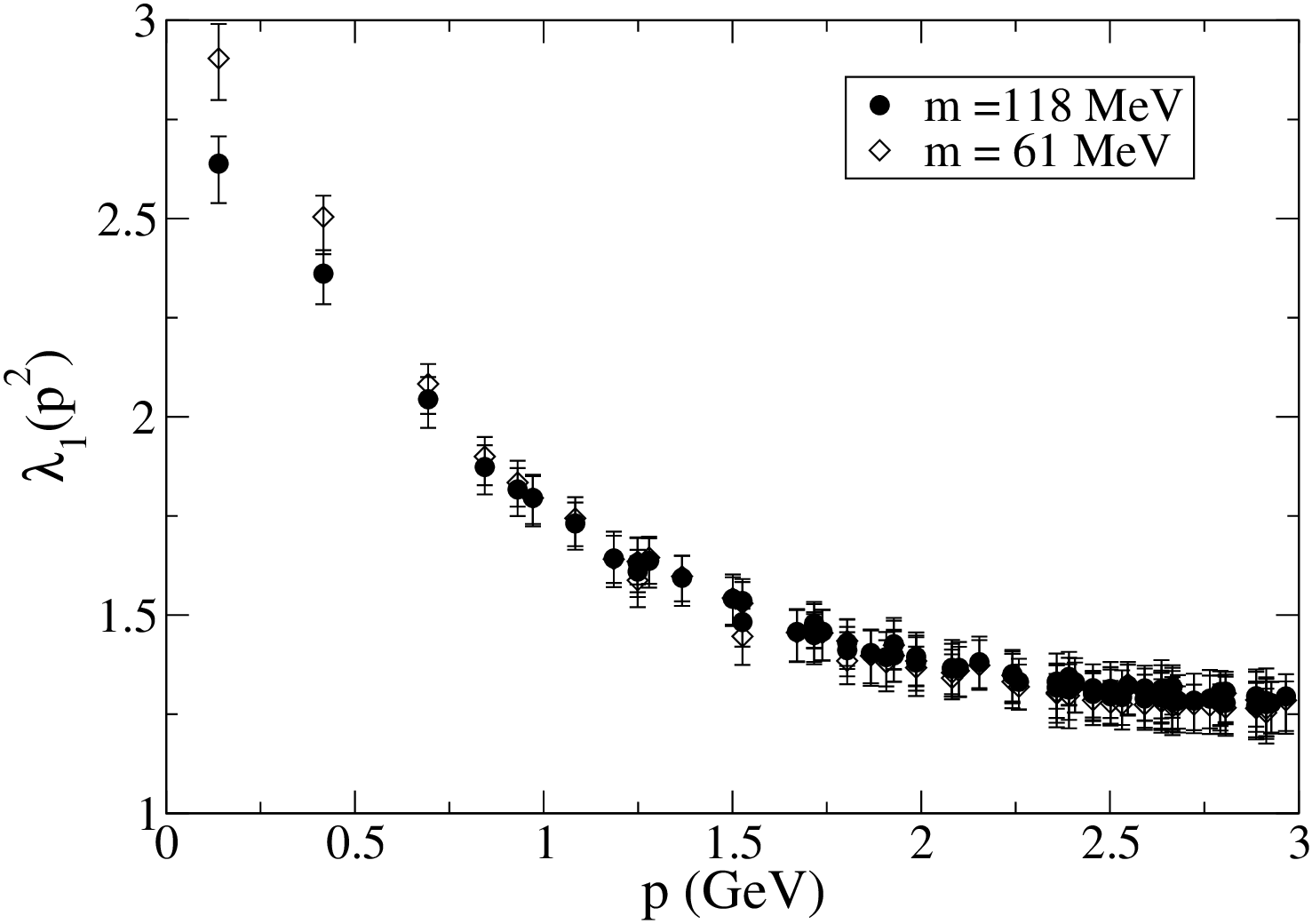}
\includegraphics*[width=0.49\textwidth]{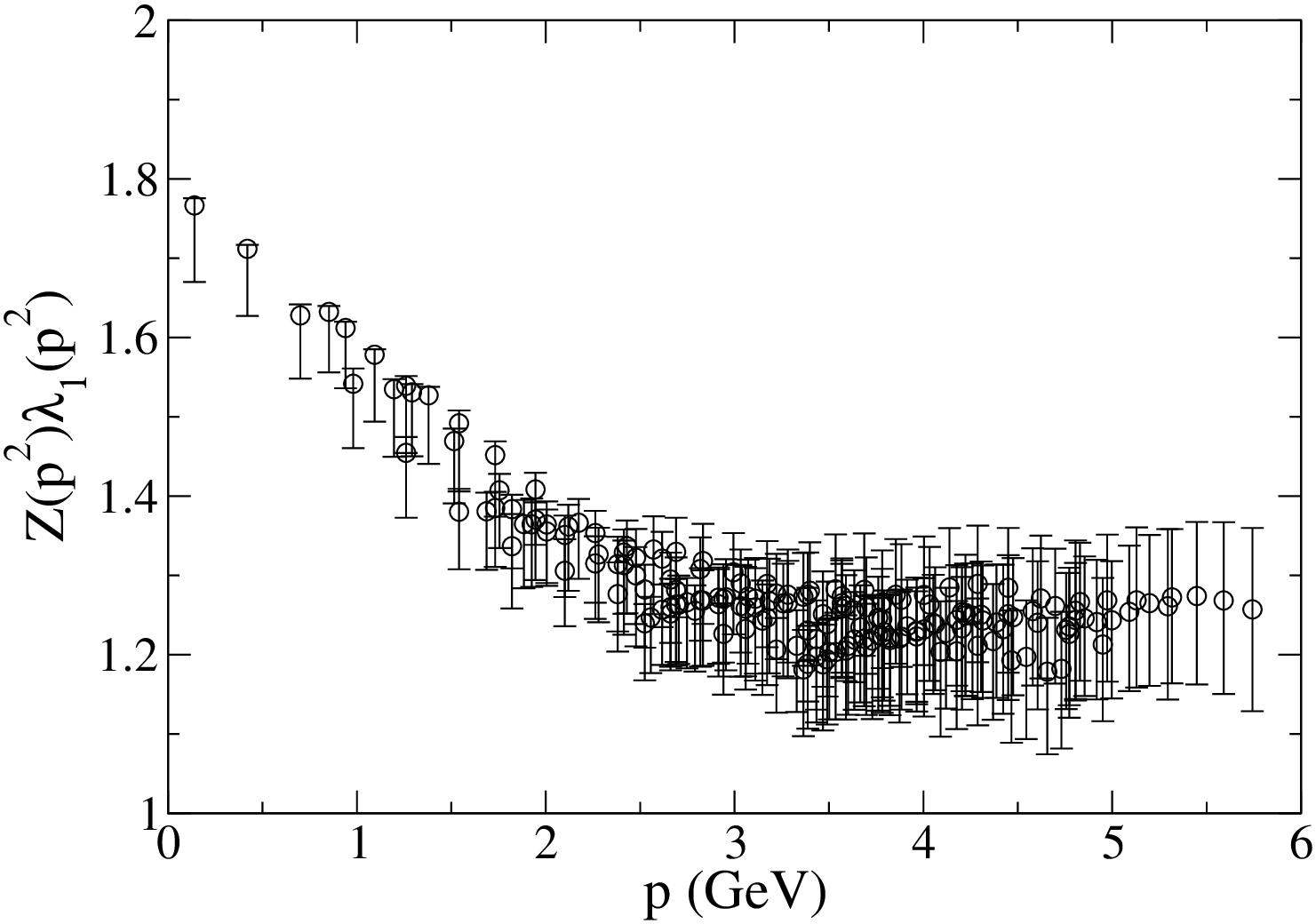}
\caption{Left: The unrenormalised form factor $\lambda_1(p^2)$ as a
function of $p$.  Right: $\lambda_1(p^2)$ multiplied by the quark
renormalisation function $Z(p)$, for $m=118$ MeV.  In an abelian
theory, this would be a constant.}
\label{fig:l1-asym}
\end{figure}

In Figure~\ref{fig:l1-asym} we show $\lambda_1$ at $q=0$ for our two quark
masses.  We see a clear enhancement of the form factor in the
infrared, and virtually no dependence on the quark mass beyond the
three most infrared points.  In the right-hand panel we show the
deviation of the non-abelian form factor from the abelian expression
in (\ref{eq:bc}).  Although much of the infrared enhancement is seen
to be present in the abelian case, we also find a definite deviation
of 50\% for the most infrared points.  However, the slight mass
dependence in the infrared can be fully accounted for by the abelian
behaviour.

\begin{figure}[ht]
\includegraphics*[width=0.49\textwidth]{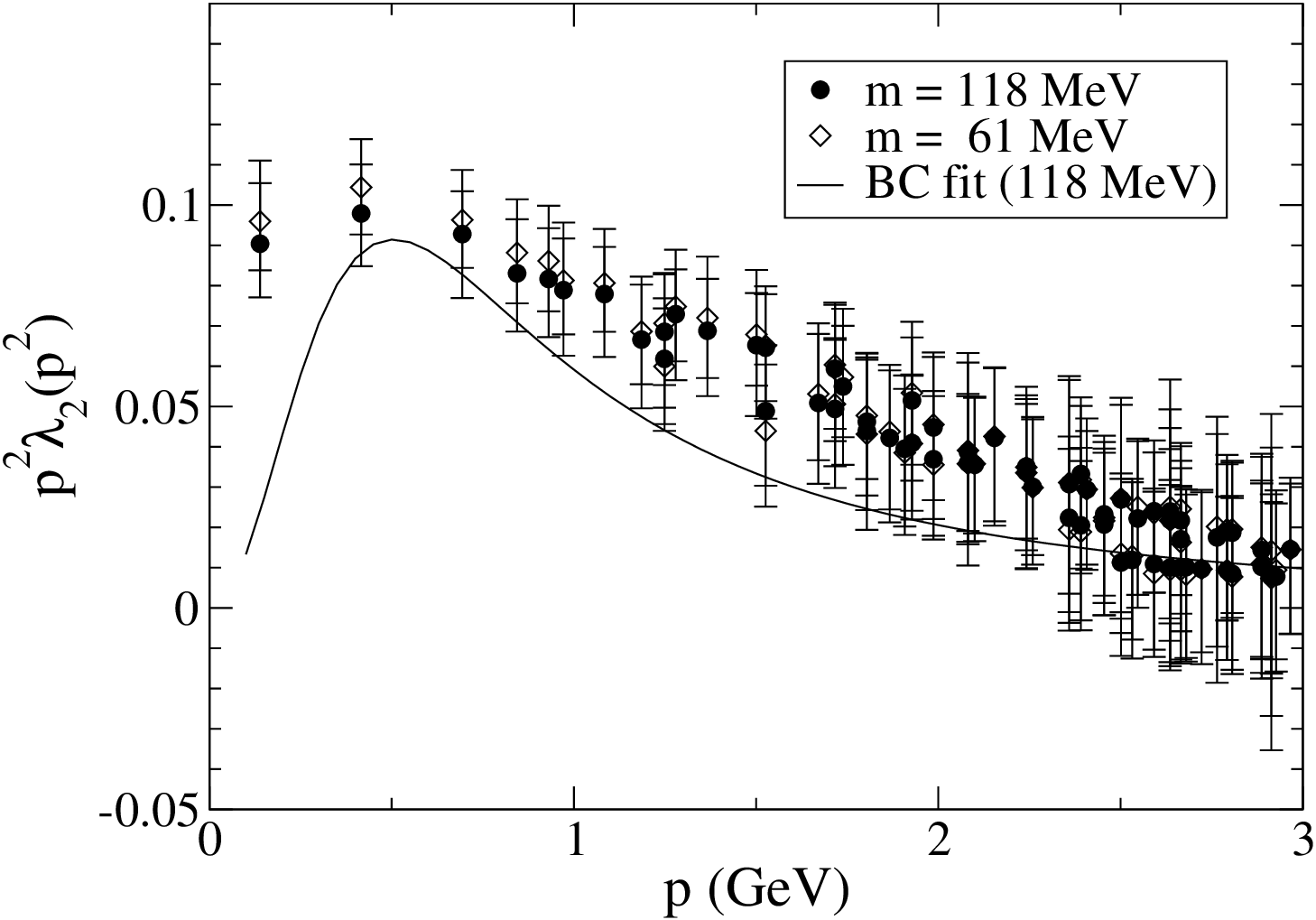}
\includegraphics*[width=0.49\textwidth]{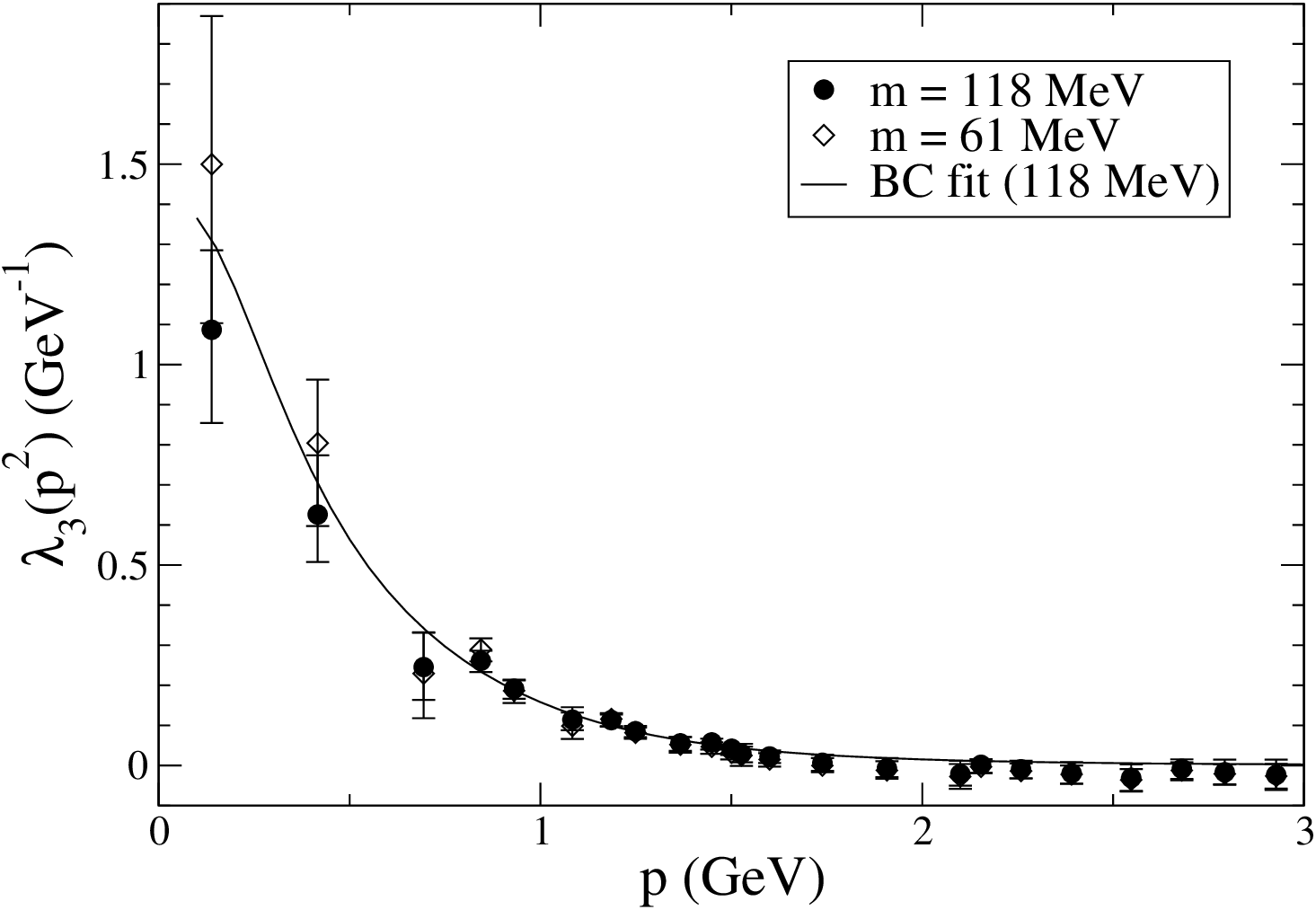}
\caption{The unrenormalised form factors $p^2\lambda_2(p^2)$ (left)
and $\lambda_3(p^2)$ (right) as functions of $p$.  Also shown are
the abelian (Ball--Chiu) forms of (\protect\ref{eq:bc}), for
the heavier mass.}
\label{fig:l23}
\end{figure}

In Figure~\ref{fig:l23}, we show the form factors $p^2\lambda_2$ and $\lambda_3$
as functions of $p$, where we have 
subtracted off the lattice tree-level behaviour to bring them closer
to the continuum.  At large momenta this procedure implies large
cancellations, and our results therefore contain little 
information beyond $p\sim3$ GeV.
The abelian form (\ref{eq:bc}) has been determined by fitting the
quark propagator form factors $Z=1/A$ and $M=B/A$ to functional
forms\cite{Bowman:2002bm}.  Our results for $\lambda_2$ appear to be
quite consistent with the abelian form, except for the most infrared
point where finite volume errors may be substantial.  For $\lambda_3$,
our results match the abelian form almost perfectly, indicating that
the non-abelian contribution to this form factor may be negligible.
In both cases, we find again at most a slight mass dependence.

\section{Outlook}
\label{sec:outlook}

In addition to the form factors presented here, we are currently
studying the chromomagnetic moment form factor ($\tau_5$ in the
notation of Ref.\cqqg) at $q=-2p$.  These results will be presented in
a forthcoming paper.  We are also planning to extend our study to the
full kinematical space available.  In the longer term, simulations
using fermion actions with improved tree-level behaviour, and on
larger volumes, will be performed in order to reduce systematic
errors.



\begin{thebibliography}{1}

\bibitem{Skullerud:2002ge}
J.~Skullerud and A.~K{\i}z{\i}lers{\"u},
\newblock JHEP {\bf 09}, 013 (2002), [hep-ph/0205318].

\bibitem{Ball:1980ay}
J.~S. Ball and T.-W. Chiu,
\newblock Phys. Rev. {\bf D22}, 2542 (1980).

\bibitem{Bowman:2002bm}
P.~O. Bowman, U.~M. Heller and A.~G. Williams,
\newblock Phys. Rev. {\bf D66}, 014505 (2002), [hep-lat/0203001].

\end{thebibliography}

\end{document}